\documentstyle[aps,preprint]{revtex}
\begin{document}
\tightenlines
\draft
\title{Ultralight particle creation during Fresh inflation}
\author{Mauricio Bellini\footnote{E-mail address: mbellini@mdp.edu.ar;
bellini@ginette.ifm.umich.mx}}
\address{Instituto de F\'{\i}sica y Matem\'aticas, \\
Universidad Michoacana de San Nicol\'as de Hidalgo, \\
AP:2-82, (58041) Morelia, Michoac\'an, M\'exico.}
\maketitle
\begin{abstract}
I study ultralight particle creation 
which becomes from the Yukawa interaction between the
inflaton and the thermal bath during fresh inflation.
Particle creation is important in the first
stages of fresh inflation, when the nonequilibrium thermal
effects are important.
I find that the number
density of the ultralight created particles
is more important as the scale factor growth rate is more large
(i.e., for $p$ large --- $a \sim t^p$). Ultralight boson fields
created during fresh inflation could be an alternative mechanism
to cosmological constant to explain the discrepancy between the
observed $\Omega_m \simeq 0.2$ and $\Omega_{tot}\simeq 1$, predicted
by inflationary models.
\end{abstract}
\vskip 2cm
\noindent
{\rm Pacs:} 98.80.Cq \\
\vskip 2cm
\section{Introduction}

Recently a new model of inflation called {\em fresh inflation}
was introduced\cite{1}. 
In this model initially the universe
there is not thermalized
[$\rho_r(t=t_0)=0$], and
begins from an unstable primordial matter field
perturbation with energy density nearly $M^4_p$ and chaotic
initial conditions. 
Later, it describes a second order
phase transition, and the inflaton rolls down towards its minimum
energetic configuration.
Particles production and heating occur together during the rapid
expansion of the universe, so that the radiation energy density
grows during fresh inflation ($\dot\rho_r >0$). The Yukawa interaction
between the inflaton field and other scalar fields in a thermal bath
($\delta = \Gamma(\theta) \dot\phi^2$),
lead to dissipation which is responsible for the slow rolling of the
inflaton field. So, the slow-roll conditions are physically justified
and there are not a requirement of a nearly flat potential in fresh
inflation. 
As the viscosity is large enough at the end of fresh inflation,
the inflaton will reach a slow-roll regime due to its dynamics becomes
overdamped ($\Gamma \gg H$).
This fact also provides thermal equilibrium
in the last phase of fresh inflation.

In this work I discuss creation of ultralight 
scalar fields which underwent in the fresh inflationary scenario\cite{1}.
Ultralight scalars have been previously discussed in the context
of pseudo-Nambu-Goldstone bosons\cite{2}. The basic idea is that
as the inflaton relaxes toward its minimum energy configuration, it
will decay into lighter fields, generating an effective viscosity.
That this indeed happens in warm inflation \cite{2a,2aa,2aaa}, and
also has been demonstrated in detail in Refs. \cite{2b}.

I consider a spatially globally
flat Friedmann - Robertson - Walker (FRW) metric
\begin{equation}
ds^2 = -dt^2 + a^2(t) dx^2.
\end{equation}
Here, $a$ is the scale factor of the universe. Furthermore,
$H=\dot a/a$ is the Hubble parameter.
The Lagrangian for a $\phi$-scalar field minimally coupled
to gravity, which also interacts with another $\psi$-scalar field
by means of a Yukawa interaction,
\begin{equation}\label{1}
{\cal L} = - \sqrt{-g} \left[\frac{R}{16\pi G} +\frac{1}{2}
g^{\mu\nu} \phi_{,\mu}\phi_{,\nu} + V(\phi)+
{\cal L}_{int}\right] ,
\end{equation}
where $g^{\mu\nu}$ is the metric
tensor, $g$ is its determinant and $R$ is the scalar curvature. 
The interaction Lagrangian is given by 
${\cal L}_{int} \sim -{\rm g}^2 \phi^2 \psi^2$, where $\psi$ is
a scalar field in the thermal bath.
Furthermore,
the indices $\mu,\nu$ take the values $0,..,3$
and the gravitational constant is $G=M^{-2}_p$ (where
$M_p = 1.2 \times 10^{19} \  GeV$ is the Planckian mass).

The dynamics for the spatially homogeneous inflaton field is given
by
\begin{equation}\label{sf}
\ddot\phi + \left(3 H + \Gamma \right) \dot\phi  +V'(\phi) =0,
\end{equation}
where 
$V'(\phi) \equiv {dV\over d\phi}$, $V(\phi) = [{\cal M}^2(0)/2] \phi^2 +
[\lambda^2/4]\phi^4$ (where ${\cal M}^2(0) >0$ is the squared mass at
zero temperature) and
$H(\phi) = 4 \sqrt{\pi G/[3(4-3F)]} {\cal M}(0) \phi$. For this
model (the reader can see a detailed development of fresh inflation
in Ref. \cite{1}), the scale factor of the universe evolves as $a \sim
t^{2/(3F)}$, where $F= - 2\dot H/(3H^2)$ is considered as constant.
The term $\Gamma \dot\phi$ is added in the scalar field equation
of motion (\ref{sf}) to describe the continuous energy transfered from
$\phi$ to the radiation field,
where $\Gamma(\theta) = g^4_{eff}/(192 \pi) \theta$ ($\theta$ is the
temperature)
and $\theta(t)$ evolves as\cite{1}
\begin{equation}
\theta(t) = \frac{192 \pi}{g^4_{eff} \lambda^2} \left\{ {\cal M}^2(0) \lambda^2
t + t^{-1} \left[ \lambda^2 \frac{(9F^2 -18 F +8)}{(4-3F)^2} +
{\cal M}^2(0) \pi G \frac{(192 F^2 - 72F^3 -96)}{(4-3F)^2}\right]\right\},
\end{equation}
being $g_{eff} \simeq 100$ the effective number of degrees of freedom
of the created particles.
The persistent thermal contact during
warm inflation is so finely adjusted that the scalar field evolves always
in a damped regime.

\section{Ultralight particle creation}

The fluctuations of the inflaton field $\delta\phi(\vec x,t)$ are
given by the equation of motion
\begin{equation}\label{e}
\ddot\delta\phi - \frac{1}{a^2} \nabla^2 \delta\phi + \left(3H
+ \Gamma\right)
\dot\delta\phi +  V''(\phi) \  \delta\phi =0.
\end{equation}
Here, the additional second term appears because the fluctuations
$\delta\phi$
are spatially inhomogeneous. 
Since $(a/a_0)= \left(t/t_0\right)^p$, the equation
for the modes of $\delta \phi =a^{-3/2} e^{-{1\over 2}\int \Gamma dt} \chi$,
with $p=2/(3F)$, can be written as a Klein - Gordon equation with
a time dependent parameter of mass $\mu^2(t) = {k^2_0 \over a^2}$, such
that
\begin{equation}\label{k^2}
k^2_0 = a^2 \left[\frac{9}{4}\left(H+\Gamma/3\right)^2 +
3\left(\dot H+\dot\Gamma/3\right)- V''[\phi(t)]\right].
\end{equation}
The time-dependent wave number $k_0(t)$ separates the infrared
(IR) and ultraviolet (UV) sectors.
The redefined field fluctuations can be written as a Fourier
expansion
\begin{equation}
\chi(\vec x,t) = \frac{1}{(2\pi)^{3/2}} {\Large\int} d^3k
\left[a_k \  \chi_k(\vec x,t)
+ a^{\dagger}_k \chi^*_k(\vec x,t)\right],
\end{equation}
where $\chi_k(\vec x,t) = \xi_k(t) e^{i\vec k.\vec x}$,
$\chi^{*}_k(\vec x,t)=
e^{-i\vec k.\vec x} \xi^*_k(t)$,
and ($a_k$, $a^{\dagger}_k$) are respectively the annihilation and
creation operators. The equation of motion for $\xi_k(t)$ is
\begin{equation}\label{xi}
\ddot\xi_k + \left[\frac{k^2}{a^2} - \frac{k^2_0}{a^2}\right]\xi_k=0,
\end{equation}
where $k^2_0$ is given by eq. (\ref{k^2}).
When inflation starts $\Gamma(t=t_0) \simeq 0$.
Furthermore the time derivative of the width decay $\dot\Gamma$
is nearly constant ($\dot\Gamma(t) \simeq
{\cal M}^2(0)$), so that the equation (\ref{xi}) 
can be approximated to
\begin{equation}\label{eq}
\ddot\xi^{(0)}_k +\left[\frac{k^2 t^{-2p}}{a^2_0 t^{-2p}_0}
- \left(\frac{9}{4}p^2 -3p -3\right) t^{-2}\right]\xi^{(0)}_k=0.
\end{equation}

The general solution (for $\nu \neq 0,1,2,...$)
for this equation is 
\begin{equation}\label{in}
\xi^{(0)}_k(t) = C_1 \sqrt{\frac{t}{t_0}} {\cal J}_{\nu}\left[
\frac{k t^{1-p}}{a_0 t^{-p}_0(p-1)}\right]+ C_2
\sqrt{\frac{t}{t_0}} {\cal J}_{-\nu}\left[
\frac{k t^{1-p}}{a_0 t^{-p}_0(p-1)}\right],
\end{equation}
where $\nu = {\sqrt{9p^2 -12p-11} \over 2(p-1)}$, which tends to
$3/2$ as $p \rightarrow \infty$ (i.e., $\nu \simeq 3/2$ for
$p \gg 1$), and $({\cal J}_{\nu}, {\cal J}_{-\nu})$ are the
linearly independent Bessel functions.
For large values of $p$ the
Bessel functions are ${\cal J}_{3/2}[x] \simeq
\sqrt{{ 2 \over \pi x}} \left({{\rm sin}[x] \over x} -
{\rm cos}[x]\right)$ and
${\cal J}_{-3/2}[x] \simeq \sqrt{{ 2 \over \pi x}}
\left({{\rm cos}[x]\over x} + {\rm sin}[x]\right)$.
We can take the Bunch-Davis vacuum such that $C_1=\sqrt{\pi/2}$ and
$C_2 = -\sqrt{\pi/2}$\cite{BD}. The general solution of eq. (\ref{xi})
can be written
as $\xi_k(t) = e^{\int g(t) dt} \  \xi^{(0)}_k$, where
$g(t=t_0) =0$. Notice that $\xi^{(0)}_k$ is the solution for the
modes when the interaction is negligible ($\Gamma \propto \theta \simeq 0$).
The dependence of the Yukawa interaction is in the function $g(t)$, which
only takes into account the thermal effects. Replacing $\xi_k(t) =
e^{\int g dt} \xi^{(0)}_k$ in eq. (\ref{xi}), and
taking into account eq. (\ref{eq}), one
obtains the differential equation for $g$
\begin{equation}
g^2 + \dot g = \frac{3}{2} H \Gamma + \frac{\Gamma^2}{4},
\end{equation}
which has the solution (for $g\gg {\dot\xi^{(0)}_k \over \xi^{(0)}_k}$ )
\begin{eqnarray}
&& g(t) \simeq \nonumber \\
&&
\frac{
{\cal M}\left[-\frac{3p}{16}, \frac{1}{4},\frac{{\cal M}^2(0)}{2} t^2\right]
\left[ \frac{{\cal M}^6(0)}{8} \left(
{\cal M}^2(0) t^2 + (3 p -1)\right) \right]
+
{\cal M}\left[ \frac{1}{2}
\left( 2-\frac{3p}{2} \right), \frac{1}{4},
\frac{{\cal M}^2(0)}{2} t^2 \right]
\left[ \frac{3}{8} {\cal M}^6(0) \left( 1-p \right) \right] }{
2
\left[
\frac{{\cal M}^6(0)}{8} t \left[ c_1 {\cal W}\left[-\frac{3p}{16},\frac{1}{4},
\frac{{\cal M}^2(0)}{2} t^2\right]
+ {\cal M}\left[-\frac{3p}{16},\frac{1}{4},\frac{{\cal M}^2(0)}{2} t^2\right]
\right]\right]} \nonumber \\
& + &
\frac{c_1 {\cal W}\left[-\frac{3p}{16}, \frac{1}{4},
\frac{{\cal M}^2(0)}{2} t^2\right]
\left[ \frac{{\cal M}^6(0)}{8} \left(
{\cal M}^2(0) t^2 + (3 p -1) \right) \right]
- c_1 {\cal W}\left[\frac{1}{2}
\left(2-\frac{3}{2}p\right), \frac{1}{4}, \frac{{\cal M}^2(0)}{2} t^2\right]
\left[\frac{ {\cal M}^6(0)}{2} \right]}{
2\left[\frac{{\cal M}^6(0)}{8} t
\left[ c_1 {\cal W}\left[-\frac{3p}{16},\frac{1}{4},
\frac{{\cal M}^2(0)}{2} t^2\right] + {\cal M}
\left[-\frac{3p}{16},\frac{1}{4},\frac{{\cal M}^2(0)}{2} t^2\right]
\right]\right]},
\end{eqnarray}
where $c_1$ is such that $g(t=t_0)=0$, due to initially the thermal
effects are neglibible.
Hence, the number density
of ultralight created particles $n_{u}$ when inflation
ends is
\begin{equation}
n_{u} = \frac{a^{-3} e^{- {\cal M}^2(0) t^2}}{2\pi^2}
{\Large\int}^{k_0}_{k_p} dk  k^2 \beta^2_k,
\end{equation}
where $k_p$ is the Planckian wave number and $\beta_k$ is given by 
\begin{equation}
\beta^2_k(t) = \frac{\xi^2_k}{(\xi^{(0)}_k)^2}\simeq
e^{2{\Large\int} g(t) dt},
\end{equation}
so that $n_u$ can be written as
\begin{equation}
n_{u} \sim a^{-3} F^2[\theta(t)]  {\Large\int}^{k_0}_{k_p} dk  k^2,
\end{equation}
where $F[\theta(t)] =
e^{- \frac{{\cal M}^2(0)}{2} t^2} e^{\int g(t) dt}$.
Notice that the spectral density $\delta^{(n_u)}_k \sim k^{3/2}$ of
the created particles is scale invariant in the range $k_p < k < k_0$.
This range defines the infrared sector during inflation.
\vskip 5cm
\noindent
Fig.1: Temporal evolution of the function $F$ for $p=2$.\\
\vskip 5cm
\noindent
Fig.2: Temporal evolution of the function $F$ for $p=20$. \\

Figure 1 shows the function $F[\theta(t)]$ as a function of the
time in Planckian unities for $p=2$, $g(t_0)=0$ and
${\cal M}^2(0) = 10^{-12} \  {\rm G^{-1}}$. Can be
seen that after $t \gg 10^7 \  {\rm G}^{1/2}$
the thermal effects are unimportant and $F$ tends to zero. However,
for $t < 10^7 \  {\rm G^{1/2}}$ the thermal effects are very
important for ultralight particle creation. This period agrees
with nonequilibrium thermal effects during the first stage of fresh
inflation, which also could be responsible for baryogenesis in fresh
inflation\cite{a}. Figure 2 shows the function $F[\theta(t)]$
for $p=20$. Notice that the peak is very much larger. This implies
that the number density of the created particles grows when the rate of
expansion of the universe increses.

\section{Final Comments}

Recently, the cosmological constant ($\Lambda$) has come back
into vogue. Dynamical estimates of the mass density from galaxy clusters
suggest that $\Omega_m = 0.2 \pm 0.1$ for the matter that clusters
gravitationaly\cite{ul}. However, the inflationary scenario suggests
that $\Omega_{tot} =1$. A cosmological constant is one way to resolve
the discrepancy between $\Omega_m$ and $\Omega_{tot}$.
In this work I studied the consequences of an ultralight boson
field created during fresh inflation scenario,
which is relaxing to its vacuum stated and dominating the energy
density of the Universe. With this hypothesis, the effective vacuum energy at
any epoch will be dominated by the heaviest fields that have not yet
relaxed to their vacuum state. So, at late times, such fields must
be very light. I deal with a potential without symmetry
breaking.
I find that the number density of ultralight created particles
is important in the first
stages of fresh inflation, when the nonequilibrium thermal
effects are important. However, at the end of fresh inflation
the number density of such created particles goes to zero
because the thermal equilibrium is restored ($\Gamma \gg H$).
Hence, a low density universe could exist without a $\Omega$ problem
making fresh inflation compatible with observations indicating a low
value of $\Omega_m$\cite{ul1,ul2}.
So, ultralight boson fields created during fresh
inflation could be an alternative
mechanism to cosmological constant to explain the discrepancy between
the observed value $\Omega_m\simeq 0.2$ and $\Omega_{tot}\simeq 1$, predicted
by inflationary models of the universe.

\vskip .05cm
\centerline{{\bf ACKNOWLEDGMENTS}}
\vskip .05cm
\noindent
I would like to acknowledge CONACYT (M\'exico) and CIC of Universidad
Michoacana for financial support in the form of a research grant.

\end{document}